\documentclass[12pt]{article}
\usepackage{graphicx}
\usepackage{amsmath}
\usepackage{hhline}
\usepackage{amssymb}
\usepackage{times}
\usepackage{cite}
\usepackage{array}
\usepackage{multirow}

\usepackage{color}
\definecolor{rltred}{rgb}{0.75,0,0}
\definecolor{rltgreen}{rgb}{0,0.5,0}
\definecolor{rltblue}{rgb}{0,0,0.75}

\usepackage{thumbpdf}
\usepackage[
        colorlinks=true,
        urlcolor=rltblue,       
        filecolor=rltgreen,     
        linkcolor=rltred,       
        pdftitle={Example File for pdflatex},
        pdfauthor={H1},
        pdfsubject={Template file},
        pdfkeywords={High-Energy Physics, Particle Physics},
        pdfpagemode=None,
        bookmarksopen=true]{hyperref}
 
\newlength{\dinwidth}
\newlength{\dinmargin}
\begin{document}

\newcommand{\pom}{{I\!\!P}}
\newcommand{\reg}{{I\!\!R}}
\newcommand{\slowpi}{\pi_{\mathit{slow}}}
\newcommand{\fiidiii}{F_2^{D(3)}}
\newcommand{\fiidiiiarg}{\fiidiii\,(\beta,\,Q^2,\,x)}
\newcommand{\n}{1.19\pm 0.06 (stat.) \pm0.07 (syst.)}
\newcommand{\nz}{1.30\pm 0.08 (stat.)^{+0.08}_{-0.14} (syst.)}
\newcommand{\fiidiiiful}{F_2^{D(4)}\,(\beta,\,Q^2,\,x,\,t)}
\newcommand{\fiipom}{\tilde F_2^D}
\newcommand{\ALPHA}{1.10\pm0.03 (stat.) \pm0.04 (syst.)}
\newcommand{\ALPHAZ}{1.15\pm0.04 (stat.)^{+0.04}_{-0.07} (syst.)}
\newcommand{\fiipomarg}{\fiipom\,(\beta,\,Q^2)}
\newcommand{\pomflux}{f_{\pom / p}}
\newcommand{\nxpom}{1.19\pm 0.06 (stat.) \pm0.07 (syst.)}
\newcommand {\gapprox}
   {\raisebox{-0.7ex}{$\stackrel {\textstyle>}{\sim}$}}
\newcommand {\lapprox}
   {\raisebox{-0.7ex}{$\stackrel {\textstyle<}{\sim}$}}
\def\gsim{\,\lower.25ex\hbox{$\scriptstyle\sim$}\kern-1.30ex%
\raise 0.55ex\hbox{$\scriptstyle >$}\,}
\def\lsim{\,\lower.25ex\hbox{$\scriptstyle\sim$}\kern-1.30ex%
\raise 0.55ex\hbox{$\scriptstyle <$}\,}
\newcommand{\pomfluxarg}{f_{\pom / p}\,(x_\pom)}
\newcommand{\dsf}{\mbox{$F_2^{D(3)}$}}
\newcommand{\dsfva}{\mbox{$F_2^{D(3)}(\beta,Q^2,x_{I\!\!P})$}}
\newcommand{\dsfvb}{\mbox{$F_2^{D(3)}(\beta,Q^2,x)$}}
\newcommand{\dsfpom}{$F_2^{I\!\!P}$}
\newcommand{\gap}{\stackrel{>}{\sim}}
\newcommand{\lap}{\stackrel{<}{\sim}}
\newcommand{\fem}{$F_2^{em}$}
\newcommand{\tsnmp}{$\tilde{\sigma}_{NC}(e^{\mp})$}
\newcommand{\tsnm}{$\tilde{\sigma}_{NC}(e^-)$}
\newcommand{\tsnp}{$\tilde{\sigma}_{NC}(e^+)$}
\newcommand{\st}{$\star$}
\newcommand{\sst}{$\star \star$}
\newcommand{\ssst}{$\star \star \star$}
\newcommand{\sssst}{$\star \star \star \star$}
\newcommand{\tw}{\theta_W}
\newcommand{\sw}{\sin{\theta_W}}
\newcommand{\cw}{\cos{\theta_W}}
\newcommand{\sww}{\sin^2{\theta_W}}
\newcommand{\cww}{\cos^2{\theta_W}}
\newcommand{\trm}{m_{\perp}}
\newcommand{\trp}{p_{\perp}}
\newcommand{\trmm}{m_{\perp}^2}
\newcommand{\trpp}{p_{\perp}^2}
\newcommand{\alp}{\alpha_s}

\newcommand{\alps}{\alpha_s}
\newcommand{\sqrts}{$\sqrt{s}$}
\newcommand{\LO}{$O(\alpha_s^0)$}
\newcommand{\Oa}{$O(\alpha_s)$}
\newcommand{\Oaa}{$O(\alpha_s^2)$}
\newcommand{\PT}{p_{\perp}}
\newcommand{\JPSI}{J/\psi}
\newcommand{\sh}{\hat{s}}
\newcommand{\uh}{\hat{u}}
\newcommand{\MP}{m_{J/\psi}}
\newcommand{\PO}{I\!\!P}
\newcommand{\xbj}{x}
\newcommand{\xpom}{x_{\PO}}
\newcommand{\ttbs}{\char'134}
\newcommand{\xpomlo}{3\times10^{-4}}  
\newcommand{\xpomup}{0.05}  
\newcommand{\dgr}{^\circ}
\newcommand{\pbarnt}{\,\mbox{{\rm pb$^{-1}$}}}
\newcommand{\gev}{\,\mbox{GeV}}
\newcommand{\WBoson}{\mbox{$W$}}
\newcommand{\fbarn}{\,\mbox{{\rm fb}}}
\newcommand{\fbarnt}{\,\mbox{{\rm fb$^{-1}$}}}
\newcommand{\ee}{ e^+ e^-}
\newcommand{\ep}{ ep}
\newcommand{\xp}{ x_p}
\newcommand{\MeV}{\rm MeV}
\newcommand{\GeV}{\rm GeV}
\newcommand{\acm}{$<\!n\!>$}
%
%
\newcommand{\qsq}{\ensuremath{Q^2} }
\newcommand{\gevsq}{\ensuremath{\mathrm{GeV}^2} }
\newcommand{\et}{\ensuremath{E_t^*} }
\newcommand{\rap}{\ensuremath{\eta^*} }
\newcommand{\gp}{\ensuremath{\gamma^*}p }
\newcommand{\dsiget}{\ensuremath{{\rm d}\sigma_{ep}/{\rm d}E_t^*} }
\newcommand{\dsigrap}{\ensuremath{{\rm d}\sigma_{ep}/{\rm d}\eta^*} }
\def\Journal#1#2#3#4{{#1} {\bf #2} (#3) #4}
\def\NCA{\em Nuovo Cimento}
\def\NIM{\em Nucl. Instrum. Methods}
\def\NIMA{{\em Nucl. Instrum. Methods} {\bf A}}
\def\NPB{{\em Nucl. Phys.}   {\bf B}}
\def\PLB{{\em Phys. Lett.}   {\bf B}}
\def\PRL{\em Phys. Rev. Lett.}
\def\PRD{{\em Phys. Rev.}    {\bf D}}
\def\ZPC{{\em Z. Phys.}      {\bf C}}
\def\EJC{{\em Eur. Phys. J.} {\bf C}}
\def\CPC{\em Comp. Phys. Commun.}

\begin{titlepage}

\begin{flushleft}
{\tt DESY 07-065    \hfill    ISSN 0418-9833} \\
{\tt May  2007}                  \\
\end{flushleft}

\vspace{2cm}

\begin{center}
\begin{Large}

{\bf Charged Particle Production in High $Q^2$ Deep-Inelastic Scattering at HERA}

\vspace{2cm}

H1 Collaboration

\end{Large}
\end{center}

\vspace{2cm}

\begin{abstract}
The average charged track multiplicity and the normalised distribution of the  scaled momentum, $\xp$, of charged final state hadrons are measured in deep-inelastic $\ep$ scattering at high $Q^2$ in the Breit frame of reference. The analysis covers the range of photon virtuality $100 < Q^2 < 20\,000~\GeV^{2}$. Compared with previous results presented by HERA experiments this analysis has a significantly higher statistical precision and extends the phase space to higher $Q^{2}$ and to the full range of $\xp$. The results are compared with $\ee$ annihilation data and with various calculations based on perturbative QCD using different models of the hadronisation process.
\end{abstract}

\vspace{1.5cm}

\begin{center}
Submitted to Phys. Lett. B.
\end{center}

\end{titlepage}

%
%
%
\begin{flushleft}

F.D.~Aaron$^{5,49}$,           
A.~Aktas$^{11}$,               
C.~Alexa$^{5}$,                
V.~Andreev$^{25}$,             
B.~Antunovic$^{26}$,           
S.~Aplin$^{11}$,               
A.~Asmone$^{33}$,              
A.~Astvatsatourov$^{4}$,       
S.~Backovic$^{30}$,            
A.~Baghdasaryan$^{38}$
% ...

\bigskip{\it
% ...
$ ^{4}$ Inter-University Institute for High Energies ULB-VUB, Brussels;
          Universiteit Antwerpen, Antwerpen; Belgium$^{ c}$ \\
 $ ^{5}$ National Institute for Physics and Nuclear Engineering (NIPNE) ,
          Bucharest, Romania \\
% ...
$ ^{11}$ DESY, Hamburg, Germany \\
% ...
$ ^{25}$ Lebedev Physical Institute, Moscow, Russia$^{ e}$ \\
 $ ^{26}$ Max-Planck-Institut f\"ur Physik, M\"unchen, Germany \\
% ...
$ ^{30}$ Faculty of Science, University of Montenegro,
          Podgorica, Montenegro$^{ e}$ \\
% ...
$ ^{33}$ Dipartimento di Fisica Universit\`a di Roma Tre
          and INFN Roma~3, Roma, Italy \\
% ...
$ ^{38}$ Yerevan Physics Institute, Yerevan, Armenia \\
% ...
$ ^{49}$ Also at Faculty of Physics, University of Bucharest,
          Bucharest, Romania \\
% ...
$ ^c$ Supported by FNRS-FWO-Vlaanderen, IISN-IIKW and IWT
      and  by Interuniversity
Attraction Poles Programme,
      Belgian Science Policy \\
% ...
$ ^e$ Supported by the Deutsche Forschungsgemeinschaft \\
% ...
}
\end{flushleft}

\newpage

\section{Introduction}
\noindent
The study of parton fragmentation and hadronisation processes provides valuable insights into the non-perturbative regime of Quantum Chromodynamics (QCD). These processes may be studied at HERA using inclusive charged particle production. In deep-inelastic $\ep$ scattering (DIS) the measurement of particle momentum spectra can be performed in the current hemisphere of the Breit frame~\cite{breit}, where the photon virtuality, $Q$, can be related to the momentum of the scattered parton. The charged hadron multiplicity and the distribution of their momenta scaled by $Q/2$ are the observables which are used in this analysis to study the fragmentation process in DIS. They can be directly compared with similar observables measured in one hemisphere of the hadronic final states in $\ee$ annihilation events, where particle momenta are scaled to half of the centre-of-mass energy $E^{*}/2$.

Previous comparison of DIS with several $\ee$ experiments~\cite{emc, h1frag1, zeusfrag1, zeusfrag2} have shown, in general, good agreement between these processes at high $Q$. At lower $Q$ this agreement is observed to break down due to higher order QCD processes such as Boson Gluon Fusion (BGF) and Initial state Compton QCD (ICQCD). These processes occur as part of the hard interaction in $\ep$ scattering but not in $\ee$ annihilation. They may lead to a relative depletion of the track multiplicity in the current region of DIS interactions as shown in~\cite{miguel}. Leading order matrix element Monte-Carlo programs, which use models of the parton cascade to describe QCD processes beyond leading order, have been shown to be  able to describe the spectra in $\ep$ interactions down to low $Q$.

Compared with the previous H1 publication~\cite{h1frag1} this analysis utilises a ten times larger data sample with a better understanding of the experimental uncertainties. Results are obtained for a large range in $Q$ ($10<Q<100~\GeV$), which overlaps with several $\ee$ experiments including LEP1, and for the full range of the scaled charged hadron momenta.

\section{The H1 detector}
A full description of the H1 detector can be found elsewhere~\cite{h1detector} and only the components most relevant for this analysis are briefly mentioned here. The origin of the H1 coordinate system is the nominal $\ep$ interaction point, the direction of the proton beam defining the positive $z$--axis (forward region). 

The Liquid Argon (LAr) calorimeter measures the positions and energies of particles, including the scattered positron, over the polar angle range $4^\circ < \theta < 154^\circ$. The calorimeter consists of an electromagnetic section with lead absorbers and a hadronic section with steel absorbers. The energy resolution for electrons in the electromagnetic section is $\sigma(E)/E=11.5 \% / \sqrt{E} ~[\GeV]~\oplus 1\%$~\cite{Andrieu:1994yn}.

High $Q^{2}$ events are triggered mainly using information from the LAr calorimeter. The trigger selects localised energy deposits in the electromagnetic section. For scattered positrons with energy above $11~\GeV$ the trigger inefficiency is negligible as determined using independently triggered samples of events. There is no explicit track requirement in the trigger.

Charged particles are measured in the Central Tracking Detector (CTD) in the range \linebreak $20^\circ < \theta < 165^\circ$. The CTD comprises two large cylindrical Central Jet Chambers (CJCs) arranged concentrically around the beam-line, complemented by a silicon vertex detector~\cite{Pitzl:2000wz} covering the range $30^\circ < \theta < 150^\circ$, two $z$-drift chambers and two multiwire proportional chambers for triggering purposes, all within a solenoidal magnetic field of strength $1.16~\rm {T}$. The transverse momentum resolution is $\sigma(p_{T})/ p_{T} \simeq 0.006~p_{T}~[\GeV]~\oplus~0.02$~\cite{kleinwort2006} . In each event the tracks are used in a common fit procedure to determine the $\ep$ interaction vertex. 

\section{Data Selection}
The data used in this analysis correspond to an integrated luminosity of $44 ~\rm pb^{-1}$ and were taken by H1 in the year 2000 when protons with an energy of $920 ~\GeV$ collided with positrons with an energy of $27.5~\GeV$. 

Events are selected if the scattered positron is detected in the LAr calorimeter in the polar angular range $10^\circ < \theta_e < 150^\circ$ and with energy greater than $11~\GeV$. The kinematic phase space, calculated using the scattered positron only, is defined by requiring the photon negative momentum $Q^2$ to be in the range $100 < Q^2 < 20\,000~\GeV^{2}$ and the inelasticity $y$, defined as the fractional energy loss of the electron in the proton rest frame, to be in the range $0.05 < y < 0.6$. The polar scattering angle for a massless parton, calculated from the positron kinematics in the quark-parton model (QPM) approximation\footnote{The definition of the polar scattering angle is $\theta_{q,lab}=cos^{-1}(\frac{xs(xs-Q^{2})-4E^{2}Q^{2}}{xs(xs-Q^{2})+4E^{2}Q^{2}})$, where $E$ is the incoming positron beam energy, $s$ is the $\ep$ centre of mass energy squared and $x$ is the fraction of the proton momentum carried by the struck quark in the QPM.}, is required to be in the range $30^\circ < \theta_{q,lab} < 150^\circ$. This ensures that the current region of the Breit frame remains in the central region of the detector where there is high acceptance and track reconstruction efficiency. It should be noted that the defined kinematic phase space can be applied simply to the theoretical models. 

Additional selections are made to reduce QED radiation effects and to suppress background events. The integrated  hadronic final state is reconstructed from combined objects, built from calorimeter clusters and tracks, using an energy flow algorithm which ensures that no double counting of energy occurs.
In order to minimise the correction due to QED radiation the value of $y$ calculated from the hadronic final state using the Jacquet-Blondel method, $y_{jb}$, and that calculated from the scattered positron, $y_{e}$, are required to satisfy $y_{jb} - y_{e} > -0.15$ and $(y_{jb}-y_{e}) / y_{jb} > -0.75$. The $z$ coordinate of the event vertex is required to be within 35 cm of the nominal interaction point. This together with the rejection of events that have an event timing which does not match the HERA bunch crossing removes background from beam gas interactions. The difference between the total energy $E$ and the longitudinal component of the total momentum $P_{Z}$ , calculated from the electron and the hadronic final state, $E-P_{Z}$ is required to be in the range $35 < E-P_{Z} < 70 ~\GeV$ in order to reduce background from photoproduction events.
The event selection outlined above results in a data sample of about 60,000 events. 

The reconstructed charged tracks in the selected events are used to study the fragmentation process. Only tracks that lie within the acceptance of the CTD which are fitted to the primary vertex and have  transverse momenta above $120 ~\MeV$ are used in this analysis. In addition a variety of other track quality cuts are applied to remove badly measured tracks in a manner which can be accurately simulated. By using only tracks fitted to the event vertex the contribution from the in-flight decays of $K^{0}$'s, $\Lambda$'s, from photon conversions and from other secondary decays is minimised.

\section{Observables}

The Breit frame provides a kinematic region where the properties of the scattered quark can be studied with a well defined and relatively clean separation from the proton remnants. In the Breit frame of reference the virtual space-like photon has momentum $Q$ but no energy. The photon direction defines the negative $z'$--axis and the current hemisphere. 

Within the QPM the photon collides head on with a (massless) quark of longitudinal momentum $Q/2$. The struck quark thus scatters with an equal but opposite momentum into the current hemisphere while the proton remnants go into the opposite hemisphere. For the purpose of comparison the current region is taken to be the equivalent of one hemisphere of an $\ee$ annihilation. The energy scale, set by the virtual photon at $Q/2$, is taken to be equivalent to half of the $\ee$ centre-of-mass energy $E^{*}/2$. 

The boost to the Breit frame is defined using kinematics calculated from the properties of the scattered positron. Hadrons emerging from the interaction with negative longitudinal momenta in this frame are assigned to the current region and associated with the struck quark.

Within this analysis the average charged multiplicity, $<\!\!n\!\!>$, is defined to be the average number of charged particles in the current region of the Breit frame per event. It is compared directly with half the average charged particle event multiplicity seen in $\ee$ annihilation. 

The scaled momentum variable $\xp$ is defined to be $p_{h}/(Q/2)$ where $p_{h}$ is the momentum of a charged track in the current region of the Breit frame. In $\ee$ annihilation events the equivalent variable is $p_{h}/(E^{*}/2)$. The inclusive, event normalised, charged particle scaled momentum distribution, $D(\xp,Q)$, is calculated as $\frac{1}{N} \frac {dn} {d\xp}$, where in each $Q$ range, $N$ is the total number of selected events and $dn$ is the total number of charged tracks with scaled momentum $\xp$ in the interval $d\xp$.

\section{Phenomenology}


Fragmentation can be studied separately from the hard subprocess. The comparison of the fragmentation of the struck quark from the proton in DIS with that of a quark produced from $\ee$ annihilation allows a test of quark fragmentation universality. Results from a number of different $\ee$ experiments~\cite{ee0, ee1} at different centre of mass energies are available allowing comparison over the full $Q$ range of this analysis. 
The contribution from weakly decaying short lived particles (e.g. $K^{0}$ and $\Lambda$) is subtracted from the $\ee$ results to be consistent with our charged particle selection. This  contribution is about $8\%$ and is estimated from the $ep$ Monte Carlo program DJANGO~\cite{django} which provides in general a good description of strange particle production~\cite{zeusstrange}.


The data presented here are used to test predictions of different models of the parton cascade and hadronisation processes, implemented in various leading order matrix element Monte Carlo programs, which have been tuned to describe $\ee$ results.

The Parton Shower model (PS)~\cite{ps}, implemented in the RAPGAP~\cite{rapgap} Monte Carlo program, describes the fragmentation process as the splitting of a parent parton into two daughters ($q\to qg$, $g\to gg$, $g\to q\overline{q}$). The splitting continues, giving rise to a parton shower. The evolution of the parton shower is based on leading $\log Q^2$ DGLAP~\cite{dglap} splitting functions. The transverse momentum, $k_{T}$, of subsequently emitted partons is highly ordered. Gluon coherence, which suppresses the emission of soft gluons at wide angles, is approximately modelled by imposing angular-ordering.

In the Soft Colour Interaction model (PS+SCI)~\cite{sci} soft gluons are exchanged between the partons produced in the parton shower and the proton remnant. Soft Colour interactions are simulated using the implementation available in the LEPTO~\cite{lepto} Monte Carlo program.

In the Colour Dipole Model (CDM)~\cite{cdm}, dipoles are created between coloured partons. Gluon emission is treated as radiation from these dipoles. New dipoles are formed with the emitted gluons from which further radiation is possible. The radiation pattern of the dipoles includes interference effects, thus modelling gluon coherence. The $k_{T}$ of emitted partons are only weakly ordered, producing a picture similar to the BFKL treatment of parton evolution~\cite{bfkl}. ARIADNE~\cite{ariadne} provides an implementation of the colour dipole model and is used in the DJANGO~\cite{django}  Monte Carlo program.  

The RAPGAP, LEPTO and DJANGO Monte Carlo programs use the Lund string model of hadronisation~\cite{string} which is based on the dynamics of a relativistic string, or gluonic ``flux tube'', stretched between coloured partons. As the partons move apart they lose kinetic energy to the string creating $q\overline{q}$ pairs which form new string pieces. This process is iterated until the available energy is used up. The residual string fragments are combined into mesons and baryons.

The HERWIG Monte Carlo~\cite{herwig} program uses the parton shower model to describe the fragmentation process but incorporates the cluster model of hadronisation~\cite{cluster}. In the cluster model partons are generated in a perturbative shower. The cascade is stopped at a given cut-off, related to the minimum transverse momentum of the emitted partons. The remaining gluons are then split into light $q\overline{q}$ pairs. Coloured objects that are close to one another are combined into colourless clusters which decay isotropically in the rest frame of the original cluster into known resonances.

Next-to-leading order (NLO) perturbative QCD (pQCD) predictions based on the CYCLOPS program~\cite{cyclops}are also available. In CYCLOPS the DIS cross-section is factorised into three parts: a proton parton-density function (PDF), the full NLO matrix element (ME), and a partonic fragmentation function (FF). 
The PDF parameterisation CTEQ6M~\cite{cteq6} is taken as default and the results are cross-checked using MRST2001~\cite{mrst}. 
The factorisation and renormalisation scales are chosen as $Q$. The uncertainty arising from the scale choice is estimated by increasing and decreasing the scale by a factor of two.
Three different parameterisations of the FF are used which are obtained from NLO fits to $\ee$ data: KKP~\cite{ff1}, KRETZER~\cite{ff2}, and AKK~\cite{ff3}. 
Each parameterisation uses the appropriate quark mix (u,d,s,c,b) for $\ep$ interactions. This mix is not flavour democratic unlike the case of $\ee$ annihilations. The predictions are for light charged hadron production ($\pi^{\pm}$, $K^{\pm}$, and $^{^{_(}}\overline{p}\,^{^{_)}}$) as  measured in this analysis. 
The value of $\Lambda \frac{(5)}{MS}=266~\MeV$ is used for the PDF, ME and FF.

In order to avoid infrared singularities in the NLO calculation, the hadrons should be produced close in rapidity to the parent parton. Therefore the rapidity of the hadron in the Breit frame is required to be greater than unity~\cite{cyclopsir}. Within this analysis this corresponds to a safe limit of $x_{p}>0.1$. As a consequence the data are only compared to the NLO QCD predictions for $x_{p}>0.1$.

\section{Data Correction}

The data are corrected for detector acceptance, efficiency and resolution effects using Monte Carlo event samples generated with the RAPGAP and DJANGO programs. All generated events are passed through the full GEANT~\cite{geant} based simulation of the H1 apparatus and are reconstructed and analysed using the same programs as for the data. These Monte Carlo event samples give a good description of the data. The residual contribution of charged particles from the weak decay of neutral particles (e.g. $K^{0}$ and $\Lambda$'s) is subtracted from the data as part of the correction procedure. The effects of QED radiation are corrected for using the HERACLES~\cite{django} program incorporated within the above Monte Carlos. The total correction factor is calculated from the ratio of the number of entries in each bin at hadron level to that at detector level. The bin sizes are chosen to give high acceptance and purity\footnote{The acceptance (purity) is defined as the ratio of the number of charged hadrons generated and reconstructed in the bin to the total number of charged hadrons generated (reconstructed) in that bin.}, typically above $60\%$, with a minimum of $40\%$. The total correction factor applied to the uncorrected data points is $\sim\!\!1.1$ for $<\!\!n\!\!>$ and typically less than $1.2$ for $D(x_{p},Q)$. In general the uncertainty in the boost to the Breit frame dominates the resolution in $\xp$. The correction associated with the tracking dominates in the highest $Q^2$ region where there is a somewhat reduced acceptance for the current region of the Breit frame within the CTD.

\section{Systematic Uncertainties}
The following sources of systematic errors are considered for all measured quantities. Also presented are the resulting typical fractional error on $<\!\!n\!\!>$ and, where different, $D(x_{p},Q)$. 

\begin{itemize}
\item  The positron energy scale uncertainty is $0.7 - 3~\% $  depending on the position of the detected positron in the LAr calorimeter. This uncertainty affects both the phase space and boost calculation. Its effect is studied by repeating the analysis while varying the positron energy scale in the simulation used to correct the data. This gives an error on $<\!\!n\!\!>$ of about $2~\%$ independent of $Q$. The resulting uncertainty on $D(x_{p},Q)$ is again independent of  $Q$ but varies with $x_{p}$ from $0.5\%$ ($x_{p}\sim 0.1$) to $7\%$ ($x_{p} \sim1.0$). The scattered positron angular resolution leads to a systematic error of about $1\%$.

\item The systematic error associated with the track reconstruction efficiency is estimated to be $2.5\%$. This is applied as an independent uncorrelated error on every point and is assumed to be constant over the $p_{t}$ and $\theta$ range of the tracks. In the highest $Q$ interval an additional uncertainty due to problematic track reconstruction in dense jets with high track multiplicities leads to an extra error of $2\%$ on the average charged multiplicity and $5\%$ on $D(\xp,Q)$.

\item The hadronic energy scale uncertainty is taken to be $4\% $. Only the calculation of $E-P_{Z}$ and $y_{jb}$, used in the event selection, rely on the hadronic energy measurement. Varying the hadronic energy scale in the simulation of the samples used to correct the data gives an error of about $1\% $.

\item The uncertainty in the correction factor arising from using different Monte Carlo models in the correction procedure, taken as the full difference between correcting the data with RAPGAP or DJANGO, results in a typical error of  $1\%$ on $<\!\!n\!\!>$ and $1.5\% $ on $D(\xp,Q)$.

\end{itemize}

Apart from the error from the scattered positron energy, which is quoted separately, the individual effects of the above experimental uncertainties are combined in quadrature. The total systematic uncertainty is dominated by the systematics attributed to the tracking efficiency. In the defined kinematic region errors arising from non $\ep$ background are negligible.

\section{Results}

\begin{figure}[h] 
  \begin{center}
    \includegraphics[width=15.4cm]{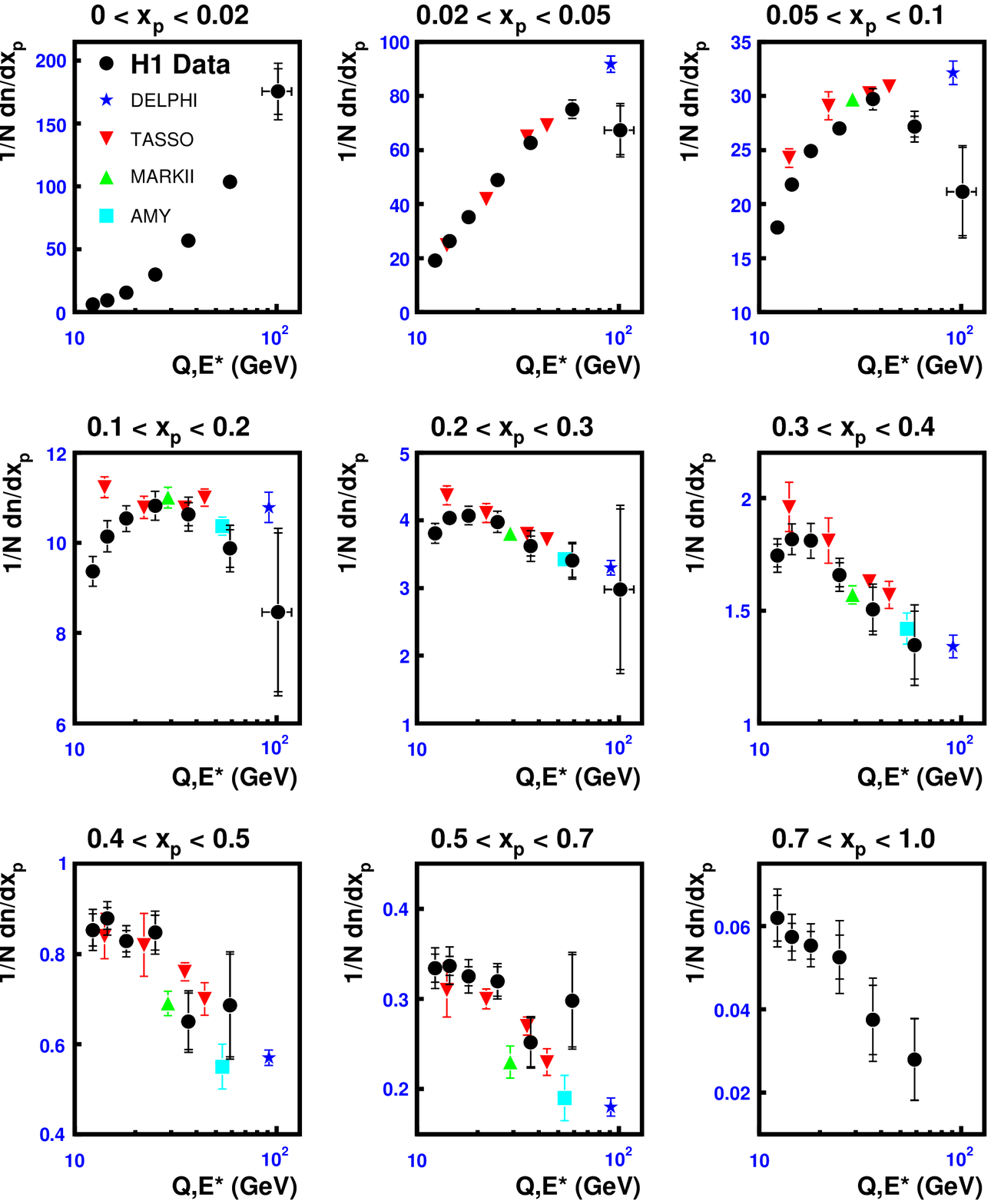}
  \end{center}
  \caption{ \label{fig:xpepvsee}  The measured normalised distributions of the scaled momentum, $\frac{1}{N} dn/dx_{p}$, as a function of $Q$ for nine different $\xp$ regions. The statistical error is shown by the inner error bar and the outer error bar represents the statistical and systematic error added in quadrature. In addition there is a further correlated error of $\sim 0.5 - 7 \%$ (increasing with $x_{p}$) coming from the electron energy scale uncertainty (not shown). The data are displayed at the average value of $Q$, the horizontal error bars represent the statistical errors given in table~\ref{table:qandx}. Data are compared to results from various $\ee$ experiments (taking $Q=E^*$). Note the suppressed zeros and large change in scale of the vertical axis moving to higher values of $\xp$. }
\end{figure}

\begin{figure}[h] 
  \begin{center}
    \includegraphics[width=16cm]{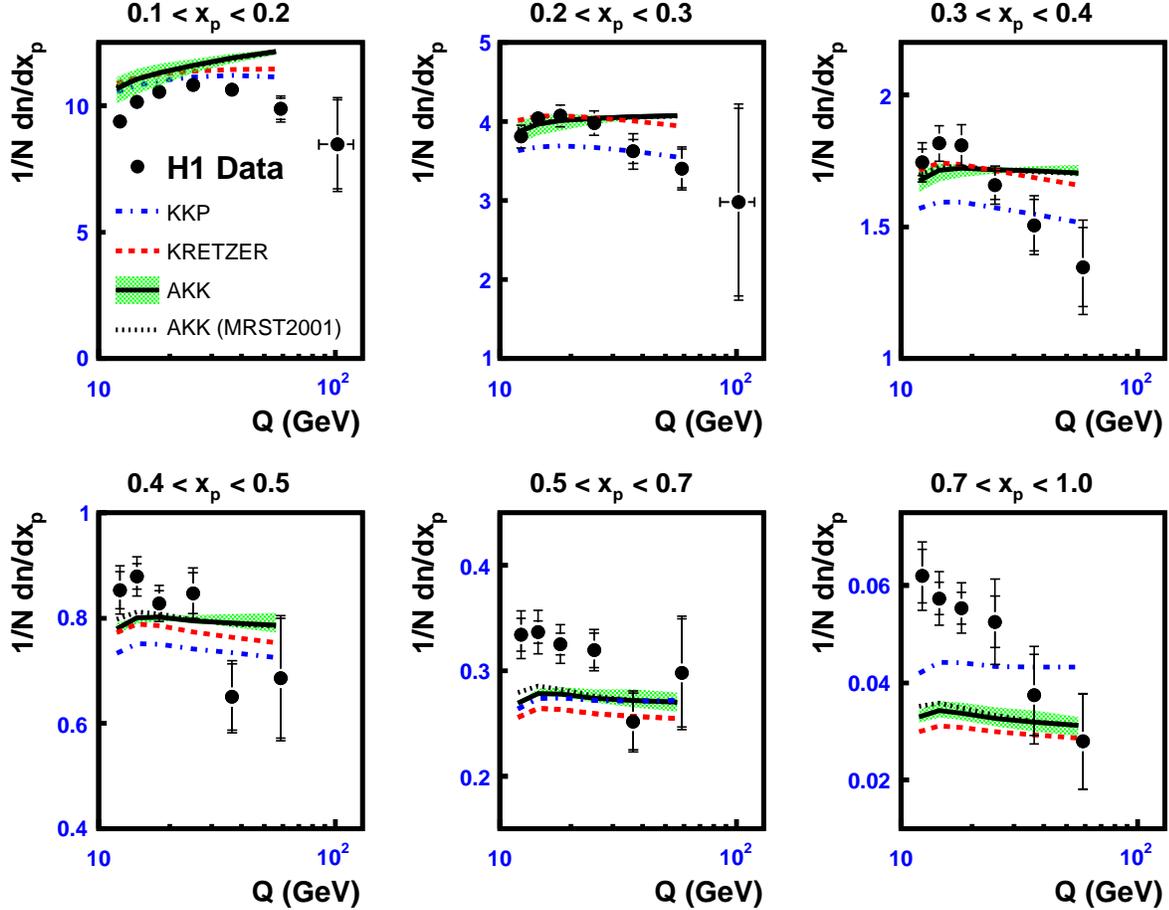}
  \end{center}
  \caption{ \label{fig:xpepvsnlo}   The measured normalised distributions of the scaled momentum, $\frac{1}{N} dn/dx_{p}$, as a function of $Q$ for six different $\xp$ regions where there exist infra red safe NLO QCD predictions. The statistical error is shown by the inner error bar and  the outer error bar represents the statistical and systematic error added in quadrature. In addition there is a further correlated error of $\sim 0.5 - 7 \%$ (increasing with $x_{p}$) coming from the electron energy scale uncertainty (not shown). The data are displayed at the average value of $Q$, the horizontal error bars represent the statistical errors given in table~\ref{table:qandx}. The data are compared to NLO QCD CYCLOPS predictions for $Q < 60~\GeV$ using three different fragmentation functions: KKP (dot-dashed line), AKK (solid), and KRETZER (dashed). The typical scale uncertainties for the AKK predictions are shown as a shaded band. As standard the CTEQ6.1 PDF is used, the effect of using the MRST2001 PDF for the AKK predictions are also shown (dotted).}
\end{figure}

The measurements of the averaged charged particle multiplicity, $<\!\!n\!\!>$, and the scaled momentum distribution, $D(\xp,Q)$, are presented. The residual  $K^{0}$ and $\Lambda$ contribution is subtracted from all results. The data are listed in tables~\ref{table:acm} and \ref{table:xp} and shown in figures \ref{fig:acm} to \ref{fig:sum}  at the average $Q$ values given in table~\ref{table:qandx}.

\subsection{Average Charged Multiplicity}

\begin{figure}[h] 
  \begin{center}
    \includegraphics[width=16cm]{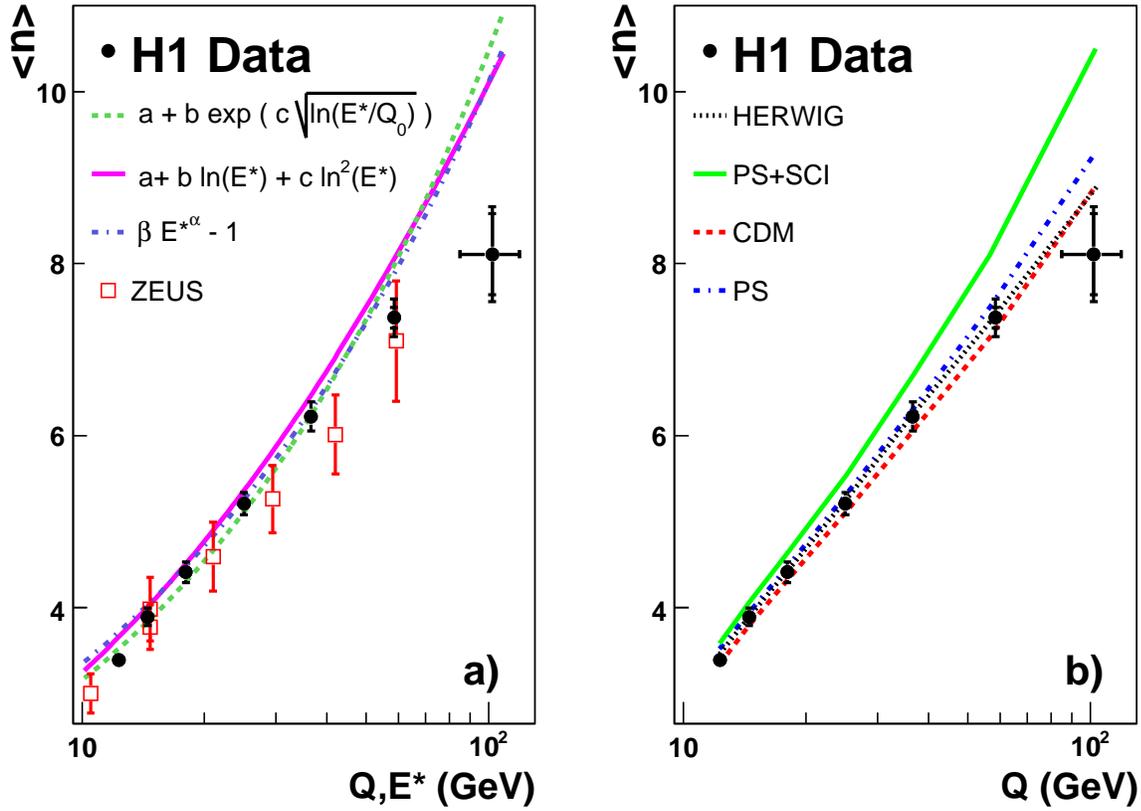}
  \end{center}
  \caption{ \label{fig:acm} The average charged multiplicity as a function of $Q$. For each measurement the statistical error is shown by the inner error bar while the outer error bar represents the statistical and systematic errors added in quadrature. In addition there is a further correlated error of $2\%$ coming from the electron energy scale uncertainty (not shown). The data are displayed at the average value of $Q$, the horizontal error bars represent the statistical errors in table~\ref{table:qandx}. The data are compared with: a) parameterisations of data from $\ee$ experiments~\cite{ee1} (taking $Q=E^*$) and with results reported by the ZEUS experiment, and b) predictions from different models of the hadronisation and parton cascade processes implemented in leading order matrix element Monte Carlo programs as described in the text.}
\end{figure}

In figure~\ref{fig:acm}a)  the measurements of $<\!\!n\!\!>$ are compared with different parameterisations  of the average charged track multiplicity per event seen in $\ee$ annihilation taken from~\cite{ee0} and with results from the ZEUS experiment~\cite{zeusfrag2}. 

The ZEUS results are in agreement, within the errors, with the H1 data. For most of the $Q$ range the H1 data are in good agreement with the parameterisation of the $\ee$ data. In the lowest $Q$ interval the data are slightly below the parameterisation of the $\ee$ data. In the highest two $Q$ intervals the measurements are clearly below the $\ee$ parameterisation.

In figure~\ref{fig:acm}b) a comparison is made with different models of the hadronisation and parton cascade processes implemented in leading order matrix element Monte Carlo programs. The models show good agreement with our data except for the one with soft colour interactions which significantly overestimates the multiplicity. 

Extensive tests have verified that the difference between the H1 data and the  $\ee$ parameterisation is not due to any sensitivity to phase space selection. At a given $Q$, $<\!\!n\!\!>$ varies by less than $2\%$ over the $x$, $y$ or $\theta_{q,lab}$ ranges. Moreover, in the vicinity of the $ Z^{0}$ resonance the admixture of heavy and light flavoured quarks in $\ep$ and $\ee$, especially with respect to the amount of $b$ quarks, gives an overall excess in the $\ee$ multiplicity of $ \sim 3\%$ which can not explain the observed difference. A possible explanation of this difference is the complexity of the DIS scattering, with extra colour connections between the scattered quark and the proton remnants, which complicates the simple analogy with $\ee$ data. The Monte Carlo models, which include some of the additional processes reflecting the complexity of the DIS interaction, are compatible with the data apart from the SCI model.

\subsection{Scaled Momenta Spectra}


In figure~\ref{fig:xpepvsee} the normalised distribution of the scaled momentum, $\frac{1}{N} \frac {dn} {d\xp}$,  is shown as a function of $Q$ for nine different intervals of $\xp$. The results are compared  to $\ee$  annihilation data~\cite{ee1}.

At low $x_{p}$ moving from low to high $Q$ there is a dramatic increase in the number of hadrons while at high $x_{p}$ the number of hadrons decreases with $Q$, i.e. the spectra becomes softer as $Q$ increases. This well known observation contradicts the scaling hypothesis of the fragmentation functions, which would imply an independence of $Q$ of the scaled momenta of hadrons. 
The $\ep$ data show similar behaviour to the $\ee$ data providing a rough demonstration of fragmentation universality. In certain phase space regions some discrepancies are visible.
It has been observed previously~\cite{h1frag1, zeusfrag1, zeusfrag2} that for $Q<10~\GeV$ there is a discrepancy between $\ep$ and $\ee$ data. This is understood in terms of higher order QCD processes depleting the current region as described in the introduction. In this analysis a similar difference is seen at low $Q$ ($Q\sim15~\GeV$) and intermediate $x_{p}$ ($0.05< \xp < 0.4$). This is reflected by the observation of a slightly reduced $<\!\!n\!\!>$  for the lowest $Q$ interval, and suggests that these higher order QCD processes have an influence in this range.
At high $Q$ ($Q>60~\GeV$) and small values of $\xp$ ($0.02< \xp < 0.2$) there are significantly less tracks observed in the $\ep$ data compared to $\ee$ annihilation. This corresponds to the observed $<\!\!n\!\!>$ values below the expectation in this $Q$ range as described in the previous section.

\begin{figure}[h] 
  \begin{center}
    \includegraphics[width=15.4cm]{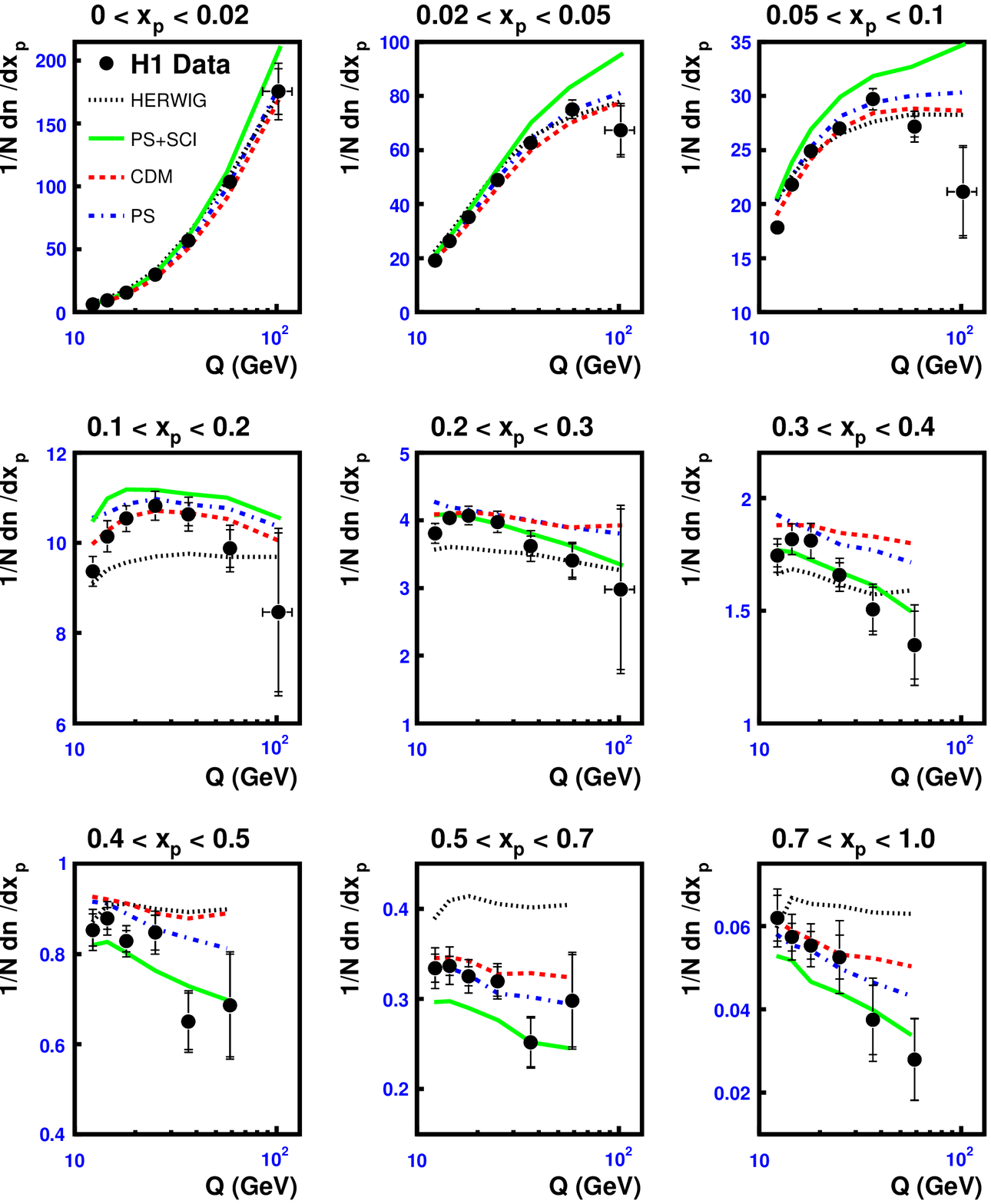}
  \end{center}
  \caption{ \label{fig:xpplotmc}  The measured normalised distributions of the scaled momentum, $\frac{1}{N} dn/dx_{p}$, as a function of $Q$ for nine different $\xp$ regions. The statistical error is shown by the inner error bar and the outer error bar represents the statistical and systematic error added in quadrature. In addition there is a further correlated error of $\sim 0.5 - 7 \%$ (increasing with $x_{p}$) coming from the electron energy scale uncertainty (not shown). The data are displayed at the average value of $Q$, the horizontal error bars represent the statistical errors given in table~\ref{table:qandx}. The data are compared to predictions from different models of the parton cascade and hadronisation processes implemented in leading order matrix element Monte Carlo programs as described in the text.}
\end{figure}

In order to investigate the fragmentation process in greater detail it is necessary to go beyond a simple comparison with $\ee$ data and compare with models that account for the processes specific to $\ep$ scattering. In figure~\ref{fig:xpplotmc} the data are compared with Monte Carlos model predictions that implement different models to describe the parton cascade and the hadronisation process. 

It can be seen from figure~\ref{fig:xpplotmc} that both CDM and the PS model provide an acceptable description of the data, with the CDM model predicting a slightly harder spectrum. Both models tend to overestimate the multiplicity at higher $Q$. The SCI model predicts a much softer spectrum than the other two models and is disfavoured by the data. This could be due to the additional gluon interactions in the SCI model which soften the spectra of partons produced by the parton shower.

The data clearly show a preference for predictions where the Lund string model of hadronisation is used (PS and CDM). HERWIG (cluster hadronisation) predicts a spectrum that is too hard compared to the data. At high $\xp$ HERWIG even fails to reproduce the observed scaling violation and predicts a flat spectrum in $Q$.

In figure~\ref{fig:xpepvsnlo} the data are compared with the predictions obtained from the NLO Monte Carlo program CYCLOPS~\cite{cyclops} for three different parameterisations of the FF in the infrared safe region as defined in section 5. The uncertainties associated with the change in scale or PDF are significantly smaller than the differences between these parameterisations. The different fragmentation function parameterisations give different results but it is evident that none of them can describe the scaling violations seen in the data. This prevents a reliable extraction of either $\alpha_{s}$ or of the fragmentation function from the data measurements using this NLO calculation.
It is interesting to note that the KKP and KRETZER  fits use different assumptions for the light quark flavour contributions to the fragmentation function while the  AKK fit uses recent $\ee$ data which include light quark tagging probabilities to constrain the strangeness contribution. The AKK and KRETZER  parameterisations are seen to agree quite well with each other.

\begin{figure}[h] 
  \begin{center}
    \includegraphics[width=16cm]{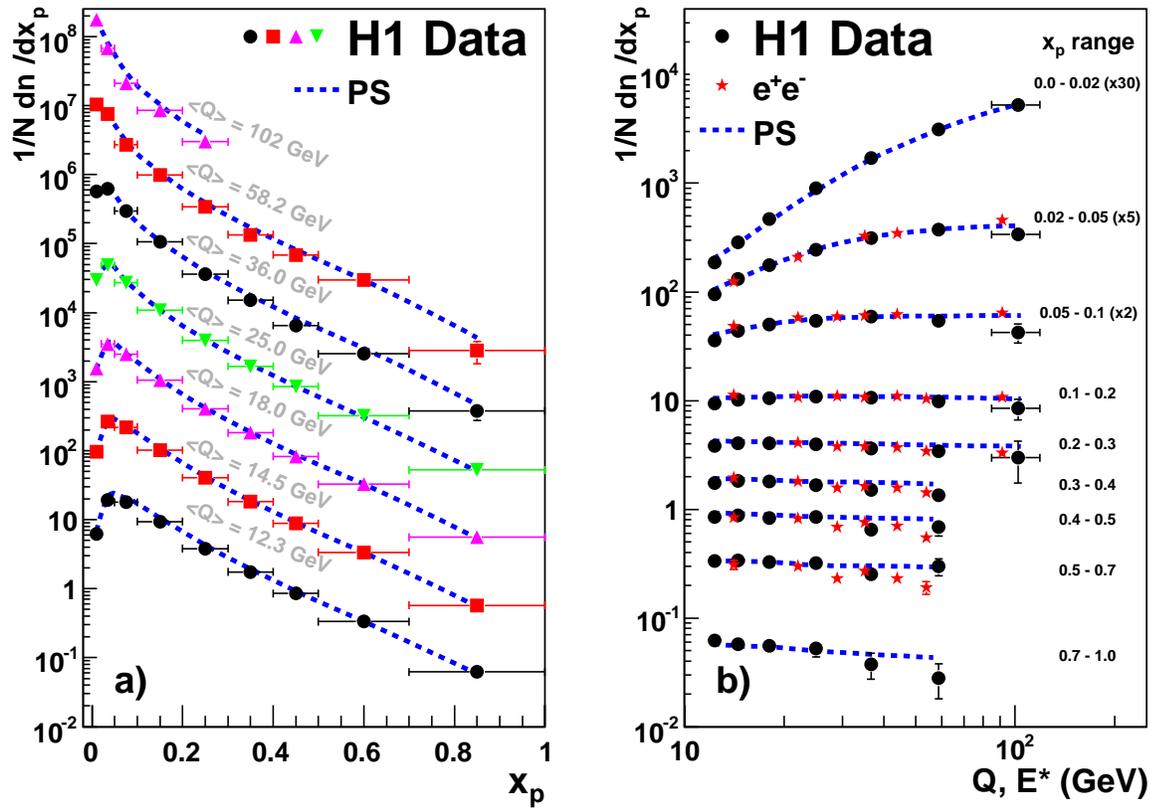}
  \end{center}
  \caption{ \label{fig:sum} The measured normalised distributions of the scaled momentum $\frac{1}{N} dn/dx_{p}$: a) as a function of $\xp$ for the different $Q$ intervals compared with the PS Monte Carlo prediction. Each Q interval, apart from the lowest, has been scaled by an additional factor of ten; and b) as a function of $Q$ for the different $\xp$ intervals compared with the $\ee$ annihilation data and the PS Monte Carlo prediction.}
\end{figure}

A summary of the results for the scaled momentum spectra is presented in figure~\ref{fig:sum}a) as a function of $\xp$ for different $Q$ intervals, where each $Q$ interval has been scaled by an additional factor ten for improved visual display, and are compared with the PS Monte Carlo prediction. In figure~\ref{fig:sum}b) the results are presented as a function of $Q$ for different $\xp$ intervals and are compared with the $\ee$ annihilation data and the PS Monte Carlo prediction.

\section{Conclusions}

The average charged multiplicity, $<\!\!n\!\!>$, and the event normalised scaled momentum distribution, $D(\xp,Q)$, of charged hadrons have been measured in $\ep$ collisions at high $Q^2$ in the Breit frame of reference and compared with $\ee$ data and a variety of models.

The results broadly support the concept of quark fragmentation universality in $\ep$ collisions and $\ee$ annihilation. A small multiplicity depletion compared to $\ee$ is observed at low $Q$ which can be attributed to higher order QCD processes occurring as part of the hard interaction in $\ep$ scattering but not in $\ee$ annihilation. At high $Q$ a large depletion is observed.
 
The best description of the data by leading order matrix element Monte Carlo programs is given by models that use the string model of hadronisation and do not include soft colour interactions. In the low and high $Q$ regions, where the comparison to $\ee$ is poor, the Monte Carlo  models are able to provide a better description of the data.

The results are compared with NLO QCD calculations as implemented in the CYCLOPS program. All three parameterisations of the fragmentation functions used in this program fail to describe the scaling violations seen in the data.

\section*{Acknowledgements}

We are grateful to the HERA machine group whose outstanding
efforts have made this experiment possible. 
We thank
the engineers and technicians for their work in constructing and
maintaining the H1 detector, our funding agencies for 
financial support, the
DESY technical staff for continual assistance
and the DESY directorate for support and for the
hospitality which they extend to the non DESY 
members of the collaboration.

We would also like to thank C. Sandoval, S. Albino and G. Kramer for providing the \linebreak CYCLOPS predictions and for useful discussions.


\newpage

\begin{table}[h]
\begin{center}
\begin{tabular}{|c||c|c|c|c|}
\hline
$Q^{2}~(\GeV^{2})$ & $<\!\!Q\!\!>~\GeV$ & $\delta Q ~\GeV$ & $<\!\!x\!\!> $ & $\delta x$ \\
\hline
\hline
$100 < Q^{2} < 175$ & 12.3 & 0.1 &  0.00370 & 0.00004 \\
\hline
$175 < Q^{2} < 250$ & 14.5 & 0.1 & 0.00952 &  0.00007 \\
\hline
$250 < Q^{2} < 450$ & 18.0 & 0.1 &  0.1559 &  0.0001 \\
\hline
$450 < Q^{2} < 1000$ & 25.0 & 0.3 &  0.0254 &  0.0003 \\
\hline
$1000 < Q^{2} < 2000$ & 36.6 & 0.8 &  0.044 & 0.001 \\
\hline
$2000 < Q^{2} < 8000$ & 58.5 & 2.1 &  0.087 &  0.003 \\
\hline
$8000 < Q^{2} < 20000$ & 102.0 & 17.0 &  0.20 & 0.03 \\
\hline
\end{tabular}
\end{center}
\caption {\label{table:qandx} Average $Q$ and $x$ values and their statistical errors for the selected events in the $Q^{2}$ intervals used in this analysis.}
\end{table}%

\begin{table}[h]
\begin{center}
\begin{tabular}{|c||c|c|c|c|}
\hline
$Q^{2}~(\GeV^{2})$ & $<\!\!n\!\!>$ & $\delta_{stat}$ & $\delta_{tot}$ & $\delta_{scale}$ \\
\hline
\hline
$100 < Q^{2} < 175$ & 3.39 & 0.02 & 0.09 & 0.05\\
\hline
$175 < Q^{2} < 250$ & 3.89 & 0.01 & 0.10 & 0.08\\
\hline
$250 < Q^{2} < 450$ & 4.41 & 0.01 & 0.12 & 0.07\\
\hline
$450 < Q^{2} < 1000$ & 5.21 & 0.03 & 0.13 & 0.07\\
\hline
$1000 < Q^{2} < 2000$ & 6.22 & 0.06 & 0.17 & 0.07\\
\hline
$2000 < Q^{2} < 8000$ & 7.37 & 0.12 & 0.22 & 0.09\\
\hline
$8000 < Q^{2} < 20000$ & 8.11 & 0.47 & 0.55 & 0.24\\
\hline
\end{tabular}
\end{center}

\caption{\label{table:acm} Average charged particle multiplicity $<\!\!n\!\!>$ as a function of $Q^{2}$ shown with the statistical error ($\delta_{stat}$), the total error including statistical and systematic errors added in quadrature ($\delta_{tot}$ ), and the correlated error coming from the electron energy scale uncertainty ($\delta_{scale}$) which is not included in the total error. }
\end{table}%

\begin{table}[htdp]
\begin{tiny}
\begin{center}
\begin{tabular}{|c||c|c|c|c|}
\hline
$Q^{2}~(\GeV^{2})$ & $\frac{1}{N} dn/dx_{p}$ & $\delta_{stat}$~[\%]  & $\delta_{tot}$~[\%]  & $\delta_{scale}$~[\%] \\
\hline
&\multicolumn{4}{c|}{ \rule{0mm}{2.5 mm} \raisebox{0.3mm}{$0.0 < x_{p} < 0.02$}} \\
\hline
$100 < Q^{2} < 175$ & 6.21 & 3.7 & 5.6 & 0.9/0.7 \\
\hline
$175 < Q^{2} < 250$ & 9.55 & 1.9 & 3.6 & 0.7/0.2 \\
\hline
$250 < Q^{2} < 450$ & 15.56 & 1.5 & 3.2 & 0.5/0.7 \\
\hline
$450 < Q^{2} < 1000$ & 29.92 & 1.7 & 3.0 & 0.6/0.7 \\
\hline
$1000 < Q^{2} < 2000$ & 56.92 & 2.3 & 3.5 & 0.3/0.6 \\
\hline
$2000 < Q^{2} < 8000$ & 103.9 & 2.3 & 3.4 & 0.2/0.7 \\
\hline
$8000 < Q^{2} < 20000$ & 175.5 & 10.3 & 12.7& 1.4/0.7 \\
\hline
&\multicolumn{4}{c|}{ \rule{0mm}{2.5 mm} \raisebox{0.3mm}{$0.02 < x_{p} < 0.05$}}\\
\hline
$100 < Q^{2} < 175$ & 19.12 & 1.6 & 3.4 & 1.5/0.9 \\
\hline
$175 < Q^{2} < 250$ & 26.29 & 1.0 & 2.8 & 1.1/1.3 \\
\hline
$250 < Q^{2} < 450$ & 35.23 & 0.8 & 2.7 & 1.1/1.3 \\
\hline
$450 < Q^{2} < 1000$ & 48.88 & 1.0 & 2.8 & 1.1/1.1 \\
\hline
$1000 < Q^{2} < 2000$ & 62.57 & 1.8 & 3.5 & 0.9/0.9 \\
\hline
$2000 < Q^{2} < 8000$ & 75.05 & 2.7 & 4.6 & 1.2/1.5 \\
\hline
$8000 < Q^{2} < 20000$ & 67.34 & 13.5 & 14.7 & 0.9/1.0 \\
\hline
&\multicolumn{4}{c|}{ \rule{0mm}{2.5 mm} \raisebox{0.3mm}{$0.05 < x_{p} < 0.1$}}\\
\hline
$100 < Q^{2} < 175$ & 17.85 & 1.3 & 2.9 & 1.1/1.4 \\
\hline
$175 < Q^{2} < 250$ & 21.81 & 0.8 & 2.7 & 1.4/1.9 \\
\hline
$250 < Q^{2} < 450$ & 24.93 & 0.7 & 2.7 & 1.2/1.5 \\
\hline
$450 < Q^{2} < 1000$ & 27.02 & 1.0 & 2.8 & 1.2/1.4 \\
\hline
$1000 < Q^{2} < 2000$ & 29.71 & 2.0 & 3.3 & 1.4/1.3 \\
\hline
$2000 < Q^{2} < 8000$ & 27.16 & 3.5 & 5.2 & 0.4/1.1 \\
\hline
$8000 < Q^{2} < 20000$ & 21.16 & 19.2 & 20.4 & 4.6/ 0.6 \\
\hline
&\multicolumn{4}{c|}{ \rule{0mm}{2.5 mm} \raisebox{0.3mm}{$0.1 < x_{p} < 0.2$}}\\
\hline
$100 < Q^{2} < 175$ & 9.37 & 1.2 & 3.5 & 1.1/0.9 \\
\hline
$175 < Q^{2} < 250$ & 10.14 & 0.8 & 3.4 & 1.8/2.1 \\
\hline
$250 < Q^{2} < 450$ & 10.54 & 0.8 & 2.7 & 1.6/1.7 \\
\hline
$450 < Q^{2} < 1000$ & 10.82 & 1.3 & 3.0 & 1.3/1.4 \\
\hline
$1000 < Q^{2} < 2000$ & 10.64 & 2.5 & 3.6 & 1.2/1.5 \\
\hline
$2000 < Q^{2} < 8000$ & 9.88 & 4.3 & 5.2 & 1.0/1.7 \\
\hline
$8000 < Q^{2} < 20000$ & 8.46 & 20.8 & 21.9 & 2.0/4.7 \\
\hline
&\multicolumn{4}{c|}{ \rule{0mm}{2.5 mm} \raisebox{0.3mm}{$0.2 < x_{p} < 0.3$}}\\
\hline
$100 < Q^{2} < 175$ & 3.81 & 2.0 & 3.9 & 1.5/1.6 \\
\hline
$175 < Q^{2} < 250$ & 4.04 & 1.3 & 2.9 & 2.0/2.1 \\
\hline
$250 < Q^{2} < 450$ & 4.07 & 2.3 & 3.9 & 2.4/2.3 \\
\hline
$450 < Q^{2} < 1000$ & 3.98 & 2.1& 3.9 & 1.9/1.6 \\
\hline
$1000 < Q^{2} < 2000$ & 3.62 & 4.2& 6.3 & 2.5/2.8 \\
\hline
$2000 < Q^{2} < 8000$ & 3.41 & 7.3 & 8.0 & 3.0/1.4 \\
\hline
$8000 < Q^{2} < 20000$ & 2.98 & 39.9 & 41.8 & 7.6/0.2 \\
\hline

&\multicolumn{4}{c|}{ \rule{0mm}{2.5 mm} \raisebox{0.3mm}{$0.3 < x_{p} < 0.4$}}\\
\hline
$100 < Q^{2} < 175$ & 1.745 & 2.9 & 4.2 & 2.7/1.6 \\
\hline
$175 < Q^{2} < 250$ & 1.817 & 1.9 & 3.7 & 3.3/3.6 \\
\hline
$250 < Q^{2} < 450$ & 1.810 & 1.9 & 4.2 & 2.7/2.9 \\
\hline
$450 < Q^{2} < 1000$ & 1.658 & 3.2& 4.4 & 3.3/2.6 \\
\hline
$1000 < Q^{2} < 2000$ & 1.506 & 6.5 & 7.5 & 1.7/1.7 \\
\hline
$2000 < Q^{2} < 8000$ & 1.346 & 11.1 & 13.2 & 1.4/6.4 \\
\hline
&\multicolumn{4}{c|}{ \rule{0mm}{2.5 mm} \raisebox{0.3mm}{$0.4 < x_{p} < 0.5$}}\\
\hline
$100 < Q^{2} < 175$ & 0.853 & 4.2 & 5.4 & 1.7/3.0 \\
\hline
$175 < Q^{2} < 250$ & 0.879 & 2.7& 4.2 & 4.7/4.3 \\
\hline
$250 < Q^{2} < 450$ & 0.828 & 2.8 & 4.1 & 4.0/3.8 \\
\hline
$450 < Q^{2} < 1000$ & 0.847 & 4.5 & 5.6 & 2.4/3.5 \\
\hline
$1000 < Q^{2} < 2000$ & 0.650 & 9.7& 10.5 & 2.9/3.1 \\
\hline
$2000 < Q^{2} < 8000$ & 0.686 & 16.6 & 17.3 & 4.1/0.0 \\
\hline
&\multicolumn{4}{c|}{ \rule{0mm}{2.5 mm} \raisebox{0.3mm}{$0.5 < x_{p} < 0.7$}}\\
\hline
$100 < Q^{2} < 175$ & 0.334 & 4.7 & 6.8 & 3.1/2.7 \\
\hline
$175 < Q^{2} < 250$ &0.337 & 3.1 & 6.1 & 5.8/6.0 \\
\hline
$250 < Q^{2} < 450$ & 0.325 & 3.2 & 5.7 & 4.9/5.0 \\
\hline
$450 < Q^{2} < 1000$ & 0.320 & 5.1 & 6.1 & 3.7/5.0 \\
\hline
$1000 < Q^{2} < 2000$ & 0.252 & 10.7 & 11.4 & 1.0/3.1 \\
\hline
$2000 < Q^{2} < 8000$ & 0.298 & 17.2 & 18.0 & 7.4/4.8 \\
\hline
&\multicolumn{4}{c|}{ \rule{0mm}{2.5 mm} \raisebox{0.3mm}{$0.7 < x_{p} < 1.0$}}\\
\hline
$100 < Q^{2} < 175$ & 0.0620 & 8.8 & 11.2 & 5.0/7.0 \\
\hline
$175 < Q^{2} < 250$ & 0.0574 & 5.8 & 9.6 & 10.5/10.7 \\
\hline
$250 < Q^{2} < 450$ & 0.0554 & 6.0 & 9.3 & 9.3/11.2 \\
\hline
$450 < Q^{2} < 1000$ & 0.0526 & 10.0 & 16.6 & 9.3/8.2 \\
\hline
$1000 < Q^{2} < 2000$ & 0.0375 & 22.2 & 26.8 & 12.3/4.0 \\
\hline
$2000 < Q^{2} < 8000$ & 0.0280 & 34.9 & 35.2 & 5.4/8.5 \\
\hline
\end{tabular}
\end{center}
\end{tiny}

\caption{\label{table:xp} The measured normalised distribution of the scaled momentum  $\frac{1}{N} dn/dx_{p}$ as a function of $Q^{2}$ for different $x_{p}$ intervals shown with the statistical error ($\delta_{stat}$), the total error including statistical and systematic errors added in quadrature ($\delta_{tot}$ ), and the correlated error coming from the electron energy scale uncertainty ($\delta_{scale}$) which is shown as two numbers ($+/-$) and is not included in the total error. }
\end{table}%

\newpage

\newpage

\newpage

\newpage

\newpage

\end{document}